\documentclass[epj,twocolumn]{webofc}

\usepackage[varg]{txfonts}   
\woctitle{The innermost regions of relativistic jets and their magnetic fields}

\begin{document}
%
\title{A new insight into the innermost jet regions: probing extreme jet variability with LOFT}
%
%

\author{I. Donnarumma\inst{1,2}\fnsep\thanks{\email{immacolata.donnarumma@iaps.inaf.it}} \and
        A. Tramacere\inst{3}\and
        S. Turriziani\inst{4}\and
        L. Costamante\inst{5}\and
        R. Campana\inst{6,7}\and
        A. De Rosa \inst{1}\and
        E. Bozzo\inst{3}
             }
\institute{INAF-IAPS, via Fosso del Cavaliere 100, 00132, Rome, Italy
\and INFN-Roma2, via della Ricerca Scientifica 1, I-00133 Roma, Italy
\and ISDC, Department of Astronomy, University of Geneva, Chemin d'Ecogia 16, 1290 Versoix, Switzerland 
\and Department of Physics, University of Rome Tor Vergata, via della Ricerca Scientifica 1, 00133 Roma, Italy
\and Department of Physics, University of Perugia, I-06123 Perugia, Italy
\and INAF/IASF-Bologna, via Gobetti 101, 40129 Bologna, Italy
\and INFN/Sezione di Bologna, via C. Berti Pichat 6, 40127 Bologna
}
\selectlanguage{english}
\abstract{Blazars are highly variable sources over timescales that can be as low as minutes. This is  the case of the High Energy Peaked BL Lac (HBL) objects showing strong variability in X-rays, which highly correlate with that of the TeV emission. The degree of this correlation is still debated, particularly when the flaring activity is followed down to very short time scales. This correlation could challenge the synchrotron-self-Compton scenario in which one relativistic electron population dominates the entire radiative output. We argue that the LOFT Large Area Detector (10 m$^2$, LAD), thanks to its unprecedented timing capability, will allow us to detect the X-ray counterpart (2-50 keV) of the very fast variability observed at TeV energies, sheding light on the nature of X-TeV connection. We will discuss the test case of PKS 2155-304, showing as it would be possible to look for any X-ray variability occurring at very short timescales, never explored so far. This will put strong constraints on the size and the location of any additional electron population in the multi-zone scenario.  Under this perspective, LOFT and the CTA observatories, planned to operate in the same time frame, will allow us to investigate in depth the connection between X-ray and TeV emissions. We also discuss the potentialities of LOFT in measuring the change in spectral curvature of the synchrotron spectra in HBLs which will make possible to directly study the mechanism of acceleration of highly energetic electrons. LOFT timing capability will be also promising in the study of Flat Spectrum Radio Quasars (FSRQs) with flux $\ge 1$ mCrab. Constraints to the location of the high energy emission will be provided by: a) temporal investigation on second timescale; b) spectral trend investigation on minute timescales. This represents a further link with CTA  because of the rapid (unexpected) TeV emission recently detected in some FSRQs.
}
\maketitle
\section{Introduction}
\label{intro}
The Large Observatory For X-ray Timing (LOFT) is one of the ESA M3 mission candidates competing for a launch planned in 2022. LOFT will measure the equation of state of neutron stars and test General Relativistic effect in the strong field regime in Galactic black holes and AGNs \cite{feroci2012}. 
However, the mission configuration will allow us to investigate a larger set of scientific targets (novae, GRBs, tidal disruption events,...) thanks to the capabilities of the two instruments aboard LOFT: the Large Area Detector (LAD,\cite{zane2012}) and the Wide Field Monitor (WFM, \cite{brandt2012}).
The LAD is a collimated experiment endowed with a 10 m$^2$ large effective area for X-ray detection and a spectral resolution approaching that of CCD-based instruments. In details, the usage of Silicon Drift Detectors and micro-channel capillary plate collimators allow us to operate mainly in the energy range 2-30 keV with an energy (timing) resolution of $<200-260$ eV (7-10ms). The WFM envisages a set of 5 units (each unit is composed of 2-orthogonal cameras), covering $1/3$ of the sky at once and 50\% of the sky available to the LAD at any time.

\begin{figure}[t!]
\centering
\includegraphics[width=0.4\textwidth]{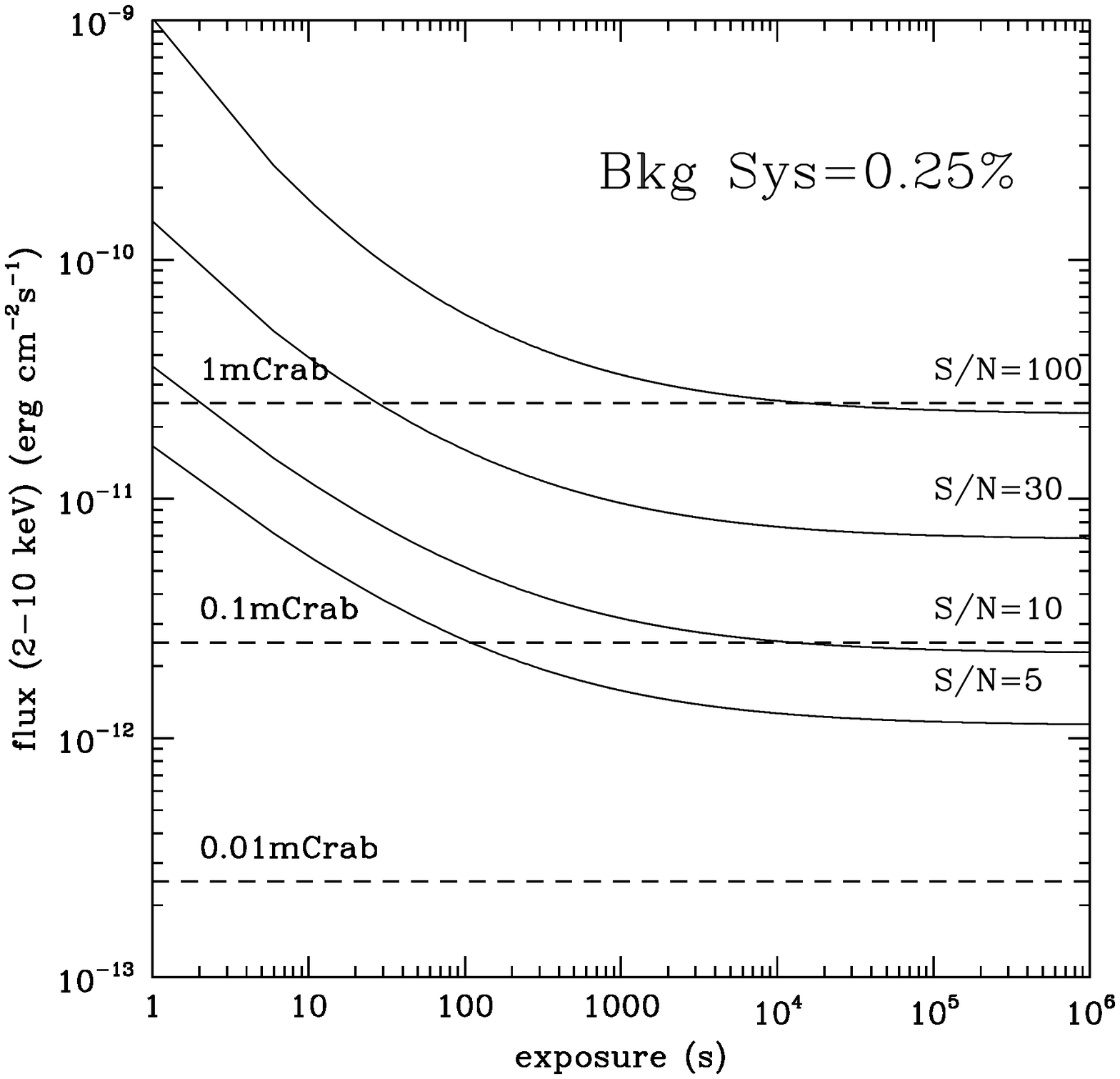}
\caption{LAD sensitivity limits at different signal to noise ratio corresponding to a systematics of 0.25\% on the background knowledge}
\label{fig:ladsens}
\end{figure}

LAD background is dominated (> 70\%) by high energy
photons of the Cosmic X-ray Background (CXB) and Earth
albedo leaking through the collimator. As these components
are relatively stable and predictable \cite{campana2013}, the anticipated LAD background systematics are estimated to be as low as $\sim 0.25\%$ even for exposure times $> 150$ ks (Figure \ref{fig:ladsens}). 

LOFT is planned in the same timeframe of other observatories which are opening up the AGN time domain on short timescales, e.g. CTA at TeV energies and SKA and ALMA in the radio band. In this context, LOFT/LAD will be able to detect a large number ($>100$) of AGNs, providing weekly monitoring (a few days for the brightest) thanks to the WFM sky coverage.

In the case of the High Energy Peaked BL Lac objects (HBLs), the X-ray band samples the synchrotron emission by the freshly accelerated and rapidly cooling highest energy electrons. In the
case of Flat Spectrum Radio Quasars (FSRQs) and Low Energy Peaked BL Lac objects (LBLs), it samples inverse Compton (IC) by the lowest-energy electrons, probing the bulk of the jet particle content and kinetic energy.
With its huge collecting area peaking at 8-10 keV and providing unprecedented observational capabilities in this energy range, the LAD will allow us to study the temporal and spectral evolution in the 2-50 keV band with unprecedented detail, for sources with fluxes above $5\times 10^{-12}$ cgs (see Figure \ref{fig:ladsens}).
The LAD characteristics will therefore match very well with those of the atmospheric Cherenkov telescopes like CTA (and HESS/MAGIC/VERITAS) because all these observatories are timing explorers, characterized by huge collecting areas in the respective bands. In the following sections, we will focus on the temporal and spectral investigation of both HBLs and FSRQs with particular emphasis on short term variability studies.

\section{High Energy Peaked BL Lac objects}
\subsection{X-ray/TeV connection}
\label{sec-hbl}

HBLs are good targets  for the LAD as its energy band (2-50 keV) probes the synchrotron emission in the region above or close to the peak where the variability is expected to be higher.

\begin{figure}
\centering
\includegraphics[width=0.4\textwidth]{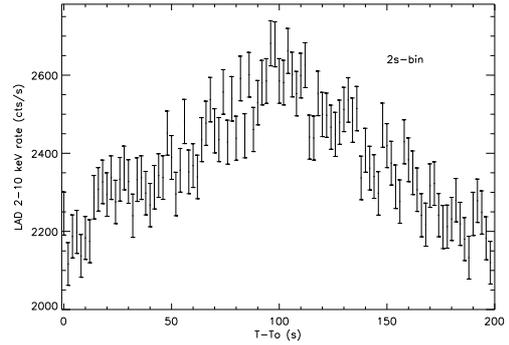}
\caption{LAD simulations of an X-ray enhancement of a 20\% over a $\sim 10$ mCrab flux in a rise time of 100 s (this is indeed the shortest timescale explored so far as in the case of PKS 2155-304). Each bin is 2 s wide.}
\label{fig:pks2155}
\end{figure}

The measurement of the TeV/X-ray lag also constrains the emission process:
e.g. synchrotron self Compton (SSC) time dependent non-homogenous modeling predicts a TeV lag equal, approximately, to the light travel time across the emission region, whereas external Compton (EC), under the assumption the main source of variability is in the relativistic electron population of the jet, predicts almost simultaneity \cite{sokolov2004}. Present data do not provide measurements of lags less than a few hundred seconds \cite{aharonian2009}, while with LAD this timescale will be pushed to a few/tens of seconds. This means that for flux of the order of $10^{-10}$ cgs in 2-10 keV (typical flaring state for many HBLs and average state for some of them) the blazar flaring activity could be followed down to the decaying tail of the light curve with comparable bin time in X-ray and TeV energy bands \cite{sol2013}.  We show in the right panel of Figure~\ref{fig:pks2155} how LAD observations will detect a flare lasting $\sim 200$ s (which is $\sim$ the shortest time bin explored so far), shaped with a flux rise of $20\%$ with respect to $10^{-10}$ cgs over 100 s. LOFT/LAD will make possible to probe even shorter timescales during the brightest peak (which means a flux by a factor of $\sim 10$ higher).
These timing capabilities will provide a \textbf{unique} diagnostics to constrain the new emerging scenario of multi-zone (multiple electron populations) SSC modeling, as suggested by the degree of correlation observed between X-rays and TeV emissions during the flaring activity \cite{aharonian2009}. The detection of flaring activity on very short timescale in X-ray and TeV will be crucial in the understanding of jet physics and its connection with the central engine: extreme variability implies very compact regions allowing us to investigate the properties of black holes and their surroundings.

Moreover, the WFM will provide an excellent trigger for CTA (and HESS/VERITAS) observations as most of the targets suitable for TeV observation are relatively bright X-ray sources.

\subsection{The acceleration mechanism}
\label{sec-hbla}
In the case of HBLs, a detailed measurement of the shape of the synchrotron spectral
component from Optical/UV to hard X-ray frequencies provides physical information about
the particle acceleration process in the jet since it directly traces the shape of the underlying particle distribution.

\begin{figure}[h!]
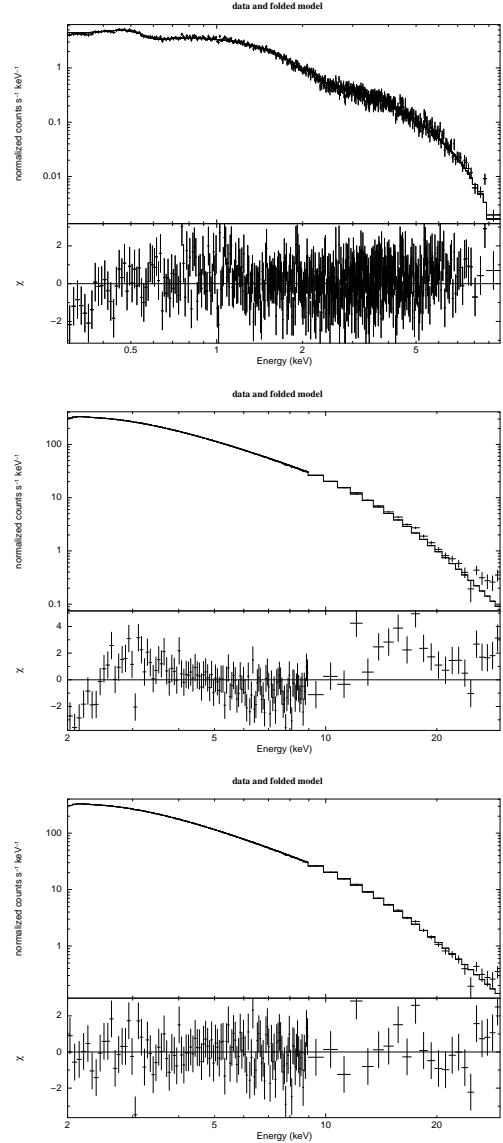

\begin{center}
\includegraphics[width=0.3\textwidth, angle=270]{LogParEpLow_FluxMed_PLC_FIT_XRT.ps}
\hspace{0.1cm}
\includegraphics[width=0.3\textwidth, angle=270]{LogParEpLow_FluxMed_PLC_FIT.ps}
\hspace{0.1cm}
\includegraphics[width=0.3\textwidth, angle=270]{LogParEpLow_FluxMed_LOG_PAR_FIT.ps}
\caption{\textsl{Top: }Swift/XRT: log-parabolic spectrum fitted with a power-law cutoff. \textsl{Center:} LAD: log-parabolic spectrum fitted with a power-law cutoff. \textsl{Bottom:} LAD: same model fitted with a log-parabola.}
\label{fig:mrk421}
\end{center}
\end{figure}

The broadband spectral distributions of several HBLs is well described by a log-parabolic
fit, with the second-degree term measuring the curvature in the spectrum. This is thought
to be a fingerprint of stochastic acceleration \cite{tramacere2007}. In this
scenario, the discrimination between a log-parabolic and power-law cut-off shape provides
a powerful tool  to disentangle acceleration dominated states, from states at the
equilibrium, or very close to, giving  solid constraints on the competition between
acceleration and cooling times, and on the magnetic field intensity. We have simulated a
pure log-parabolic spectrum peaking a 1.5 keV, resembling the typical spectrum of a HBL object in a quiescent state. We assumed a flux of $\sim 10^{-10}$ cgs in the energy range $2-20$ keV over an integration time of 10 ks. In the center and bottom panels of Figure \ref{fig:mrk421} we show the capability of the LAD to discriminate between a log-parabola (bottom panel) and a power-law cut-off model (center panel) with a high statistical significance. Thanks to the large sensitivity and the broad energy range of the LAD this result provides a relevant improvement compared to the performance of lower effective area instrument operating in the usual 0.2-10.0 keV range (simulations are performed in the case of Swift/XRT, top panel), exploring the curvature of the X-ray spectrum around 10 keV with unprecedented detail, almost a decade higher in energy than allowed by past X-ray observatories. Moreover, the capability to extract detailed spectra with a sub-ks temporal integration during the higher states, and the low energy threshold of 2 keV  will complement the current understanding provided by X-ray observatories such as NuSTAR \cite{harrison2013}.
In addition to a phenomenological log-parabolic shape, we  have simulated LAD spectra using a template numerical SED resulting from stochastic  acceleration simulations taking fully into account both acceleration and cooling processes \cite{Tramacere2011}.
We have found that, also for this more realistic scenario, the LAD can provide firm indications  on the difference between the acceleration dominated/equilibrium states. These results will be presented in a forthcoming paper (Tramacere et al. in preparation).
\section{FSRQs: X-Gamma connection and bulk-Comptonization}

Contrary to HBLs, FSRQs are blazars whose emission peaks at low energies, with the synchrotron
component peaking in the radio/IR bands and the IC one in the
Fermi and AGILE bands (50 MeV - 50 GeV). As such, they were not expected to significantly emit at VHE ($>0.2$ TeV).
It came thus as a surprise the strong and  highly variable emission
recently discovered by TeV observatories (e.g. 3C 279, 4C 21.35, PKS 1510-089).
The origin is still unclear, but the LAD can provide the answer:
if due to a tail of high-energy electrons,
their synchrotron radiation should appear in the hard X-ray band as well (like in Intermediate BL Lac or HBL objects),
dominating over the IC one and completely changing the spectral slope
(from $\Gamma \sim 1.5$ to $\Gamma>2$).  Timing is critical:
given the highly variable TeV emission (minutes-hours), such X-ray
features can disappear quickly and not be measurable by instruments of lower area or poor pointing flexibility. In this respect, LOFT/LAD larger area would provide 1-s binned light curves for flux $\ge 1$ mCrab ($\sim 2\times 10^{-11}$cgs in 2-10 keV) and, in addition, spectral trend investigation during the flaring activity will be performed on minute time scale, being the non-thermal continuum power-law constrained within a few percent.

A further diagnostics will be provided by the LAD on the jet composition,
in connection with GeV flares. If the jet is matter-dominated and ejected in blobs, the cold electrons are expected to Compton-upscatter BLR photons
yielding a spectral signature as an excess emission at $\approx(\Gamma/10)^2$ keV, where $\Gamma$ is the bulk-motion  Lorentz factor of the jet \cite{sikora2000}.
This feature, referred as bulk-Comptonization, has not been seen yet, but 1) with values of $\Gamma > 20$, this feature could actually peak around 10 keV, and thus could not be recognized so far;
2) if the jet accelerates slowly, it could be visible only for few hours, from the time the blob becomes relativistic to the time it goes outside the BLR (at a distance of $\sim10^{18}$ cm from the central engine) as reported in \cite{celotti2007} (see their Figure 1). The LAD will able to discover this feature for the first time, or put very strong upper limits on the particle content of the jet at its base.

Combining LAD repointing with radio observations with ALMA can have also a strong impact on the SED
modeling and thus on the nature of the gamma-ray emitting region,
being the higher radio frequencies (30-950 GHz) properly related to
the flaring activity.

\section{Conclusions}
We discussed the potential of LOFT for the study of blazars mainly based on its timing capabilities exploited in synergy with other observatories (radio, TeV) planned to operate in 2020s.  
On the basis of the portion of the spectral energy distribution accessible to LOFT (2-50 keV), we argue that the HBL objects are the best candidates for pointed observations with LAD. These observations will provide us the best sampling of the light curves in flaring states, given the exploration of unprecedented short timescales comparable with those achieved by the future TeV observatory CTA and in strict simultaneity. This will be possible for a large sample of TeV blazars thanks to the LAD pointing flexibility (with $\sim 70\%$ of the sky accessible). 
Therefore, LOFT will open a new window on the investigation of X-ray/TeV connection, with particular regard to: 

\begin{itemize}
\item lag measurements with at least an order of magnitude improvement over the current limit of $\sim 100$ seconds;
\item the study of the multi-zone SSC role during flares;
\item the connection between the jet variability and the central engine.
\end{itemize}

Moreover, the capability to extract detailed spectra in the range 2-50 keV with a sub-ks temporal integration (during the higher states) will allow us to study the mechanism of acceleration of highly energetic electrons, by following the curvature variations of the synchrotron emission.

Although FSRQs are less bright and variable in X-rays (see e.g. \cite{rao}) with respect to HBLs, LAD follow-up observations will allow us to investigate temporal (second timescale) and spectral variations during typical flaring activity (minute timescale). In a multifrequency context, LOFT will contribute to identify the nature and location of the high energy emission (particularly if detected at TeV energies which explore similar timescales). 

Finally, the sky coverage, the 1-day sensitivity ($\sim$ a few mCrab at 5-$\sigma$ for an on-axis source) and the pointing flexibility of the WFM as well will provide triggers for multiwavelength follow-up from radio (SKA, ALMA) up to TeV energies (CTA).






%
\bibliography{project}{}

%
%

%

%

\end{document}